\documentclass[fleqn,twoside]{article}
\usepackage{espcrc2}

\usepackage{graphicx}
\usepackage[figuresright]{rotating}

\newcommand{\AmS}{{\protect\the\textfont2
  A\kern-.1667em\lower.5ex\hbox{M}\kern-.125emS}}

\newcommand{\vev}[1]{\langle #1 \rangle}
\newcommand{\msbar}{{\overline {\rm MS}}}
\def\mdyn{m_{\rm dyn}}
\def\mval{m_{\rm val}}
\def\mres{m_{\rm res}}
\def\mhalfstrangebar{\overline{m_s}/2}
\def\bps{B_{\rm PS}}

\title{$B_K$ from Two-flavor Dynamical Domain Wall Fermions}

\author{
Taku Izubuchi\address{Institute of Theoretical Physics, 
Kanazawa University, Ishikawa 920-1192, Japan}\address[RBRC]{RIKEN-BNL Research Center, Brookhaven National Laboratory, Upton, NY 11973-5000, USA}\thanks{
This work was  partially supported by JSPS fellowship for research abroad
until February, 2003.}
for [ RBC collaboration ]\thanks{
We thank RIKEN, BNL and the US DOE for providing the facilities
essential for the completion of this work.  
Calculations have been done on the Columbia and RBRC QCDSP computers.
This work was done in
collaboration with Y. Aoki, F. Berruto, T. Blum, N. Christ, M. Creutz, 
C. Dawson, C. Kim, L. Levkova, R. Mawhinney, Y. Nemoto, J. Noaki, S. Ohta, 
K. Orginos, A. Soni and N. Yamada.  Presented  at Lattice '03, Tsukuba,
Japan.
}
}
\begin{document}

\begin{abstract}
We report preliminary results from an ongoing calculation of $B_K$ 
for $N_f=2$ dynamical QCD with domain wall fermions.  Simulations have
been done with three dynamical quark masses on $16^3 \times 32$
volumes with $L_s = 12$, where the lattice spacing is
$a^{-1} = 1.81(6)$ GeV.  
Using measurements on $\sim 70$ lattices for each dynamical mass 
and extrapolating $\mdyn=\mval$ to the kaon point,
we find $B_K^{\rm \overline{MS}}(\mu=2{\rm GeV}) = 0.503(20)$.
\end{abstract}

\maketitle

\section{INTRODUCTION}

The kaon B parameter, $B_K$, is an essential bridge between
$K_0-\overline{K_0}$ mixing experiments and the CKM matrix.  Using domain
wall fermions (DWF), quenched calculations of $B_K$
have been done \cite{RBC-WME,CPPACS-BK,Noaki_Lat03}, where operator
mixing and $O(a)$ errors are small because of the good chiral
properties of DWF.  Including dynamical fermions is an obvious
way to improve these calculations.

Using an improved HMC algorithm for DWF \cite{Izb_Lat02,CDawson_Lat03},
we have generated three ensembles of lattices of size $16^3\times 32$
for $N_f = 2$ QCD with degenerate dynamical masses of
$\mdyn=0.02,0.03,0.04$.  Our fermion action is the standard DWF action
with $L_s=12$ and $M_5=1.8$ and our gauge action is DBW2 with
$\beta=0.80$.  Extrapolating the rho mass in $\mdyn$ to the chiral
limit, we find a lattice spacing of $a^{-1} = 1.81(6)$ and a residual
mass $\mres = 0.00136(5)$.  We measure on lattices separated by 50
trajectories and have 78, 84, 57 lattices for $\mdyn = 0.02$, 0.03 and
0.04, respectively\footnote{This includes some data generated after the
conference.}.  For more details see \cite{CDawson_Lat03} and for
results with $\Delta S = 1$ matrix elements see
\cite{Mawhinney_Lat03}.

\section{CALCULATION OF $B_K$}

$B_K$ is defined by
\begin{eqnarray}
B_K \equiv 
{
\vev{\overline{K}^0 | O_{LL} | K^0}  
\over
{8\over 3}
\vev{ \overline{K}^0 | A_\mu  | 0 }
\vev{ 0 | A^\mu | K^0 }
}~~,
\label{eq:bk_def}
\end{eqnarray}
where $A_\mu=\bar s \gamma_\mu \gamma_5 d$ and $O_{LL}$ is the $\Delta
S = 2$ four-quark operator $(\bar{s}d)_{V-A} \, (\bar{s}d)_{V-A}$.  For
a general value of the mass of the pseudoscalars entering
(\ref{eq:bk_def}), we will use the symbol $\bps$, reserving $B_K$ for
the case where the pseudoscalars have the kaon mass.  

We first evaluate
this by the conventional method, where the ratio of a three-point
Green's function and two pseudoscalar--axial-vector correlators is
taken:
\begin{eqnarray}
R(t) ={
\vev{J_5^a(t_0) O_{LL}(t) J_5^a(t_1)}\over \frac{8}{3}
\vev{J_5^a(t_0) A_4^a(t)} \vev{A_4^a(t) J_5^a(t_1)}
}~~.
\end{eqnarray}
(no sum on $a$).
We use Coulomb gauge fixed wall sources at $t_0=4$ and $t_1=28$ for
valence masses, $\mval$, equal to $0.01$, 0.02, 0.03, 0.04 and 0.05.
Propagators are found with both periodic and anti-periodic temporal
boundary conditions and averaged to double the temporal length.

\begin{figure}[t]
\includegraphics[angle=270,scale=0.25]{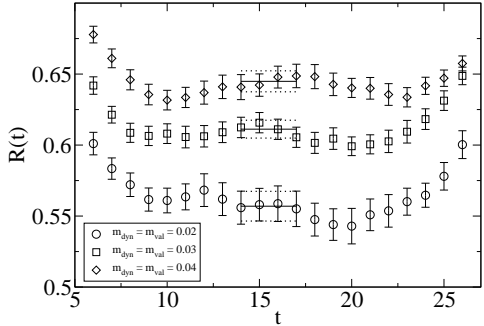}
\vspace*{-1cm}
\caption{$R(t)$ for $\mval=\mdyn = 0.02$, 0.03 and 0.04.
The lines are the average from $t = 14$ to 17.}
\label{fig:BK_CONV_plateau}
\end{figure}

In Figure \ref{fig:BK_CONV_plateau} we show $R(t)$.  We choose to fit
data from $t = 14$ to 17, same fitting range as \cite{RBC-WME}, 
to determine $\bps^{\rm lat}$.
The error bars in this figure, 
and throughout this report, assume that our
lattices are decorrelated.  The fits shown in Figure
\ref{fig:BK_CONV_plateau} give values of $\bps^{\rm lat} = 0.557(10)$,
0.611(6) and 0.645(7) for $\mdyn = \mval = 0.02$, 0.03 and 0.04
respectively.  Variations in the plateaus are noticeable at the 1-2\%
level and the assumption of 50 trajectories being sufficient for
decorrelation needs further studies and statistics.  
However, the current data certainly makes a few percent 
resolution in the lattice values achievable by 
possibly doubling the statistics.

\begin{figure}[t]
\includegraphics[angle=270,scale=0.25]{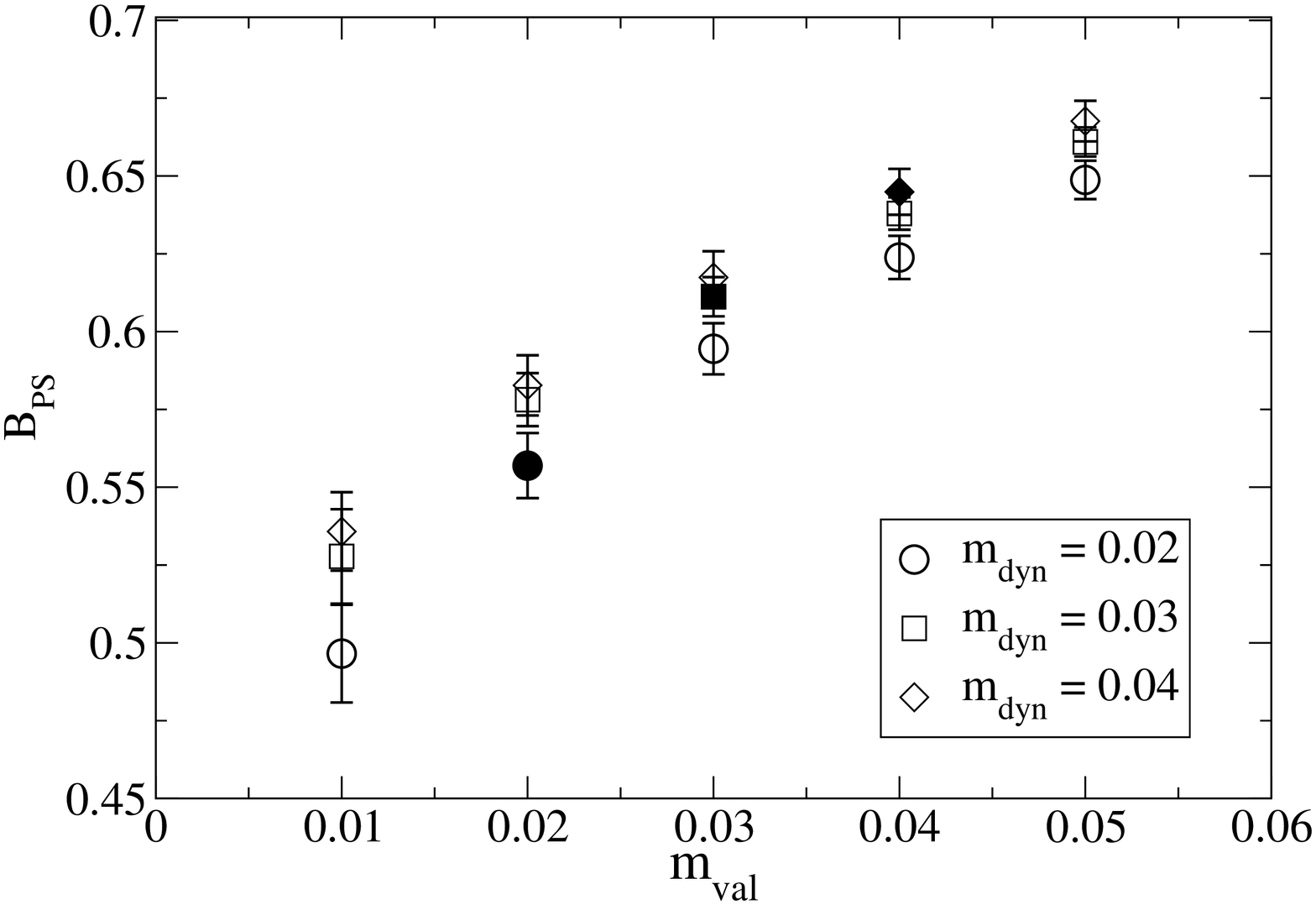}
\vspace*{-1cm}
\caption{$\bps^{\rm lat}$ for different valence and dynamical
masses.  The filled symbols have $\mval = \mdyn$.}
\label{fig:BK_CONV_vs_mass}
\end{figure}

In Figure \ref{fig:BK_CONV_vs_mass}, we show our results for
$\bps^{\rm lat}$ for degenerate valence quark masses for each different
value of $\mdyn$.  The filled points in the figure have $\mdyn = \mval$
and correspond to the fitted lines in Figure
\ref{fig:BK_CONV_plateau}.  (We also have data for non-degenerate
valence quarks and with our statistics, only the average quark mass
plays a role.) Our
simplest result to report for $B_K$ comes from
extrapolating the $\mval = \mdyn$ points in Figure
\ref{fig:BK_CONV_vs_mass} to 
$m_f = \mhalfstrangebar \equiv 0.018$. 
(A pseudoscalar with two quarks with  $m_f = \mhalfstrangebar$ 
has the mass of the kaon.)  We perform this
extrapolation using the partially quenched formula \cite{ChPTs}
\begin{eqnarray}
  \bps(m) & = & b_0 \left( 1 - {6\over (4\pi f)^2} 
  \left(M^2_{\rm PS} \log{ M^2_{\rm PS} \over
    \Lambda_{\rm \chi PT}^2 }\right)\right) \nonumber \\
  & + & b_1 M^2_{\rm PS}~~,\label{eq:ChPT}
\end{eqnarray}
fitting the three points with $\mval = \mdyn$ to determine $b_0$ and
$b_1$.  We find $b_0 = 0.267(12)$, $b_1=1.256(79)$ and $B_K^{\rm lat} =
0.541(9)$ (statistical error only), with a small $\chi^2$/d.o.f.  Note
that we are using the known coefficient of the chiral logarithm, 
$M^2_{\rm PS} = 4.06\times(m_f+m_{res}), 
f =0.0786$  and $\Lambda_{\rm \chi PT} = 1$ GeV.  The statistical error
in the lattice spacing gives an additional 2\% systematic error in the
value of $B_K^{\rm lat}$, due to the uncertainty in the strange quark
mass.

While our data makes extrapolation to $m_f = \mhalfstrangebar$ quite robust, the
value of $b_0$ is much less solid.  A small value in the chiral limit
has also been seen in the quenched calculations, but more control over
the extrapolation is needed.

We have also fit the dependence of $\bps$ on $\mval$ to (\ref{eq:ChPT})
for each fixed $\mdyn$, using the five values for $\mval$.  The
interpolated values for $B_K^{\rm lat}(\mval=\mhalfstrangebar,\mdyn)$ are
0.537(11), 0.557(9) and 0.568(10) for $\mdyn = 0.02$, 0.03 and 0.04
respectively. 
We reduce the fitting range to exclude the heaviest 
two points and see the results stay same within error.

An alternative method of extracting $B_K$ is to divide the three-point
function in the numerator of (\ref{eq:bk_def}) by a wall-wall
pseudoscalar correlator \cite{RBC-WME}.  This removes some zero mode
effects in the quenched approximation, while giving larger statistical
errors.  However here we see no difference in the central values
for results calculated using these two methods.

\section{RENORMALIZATION AND MIXING}

To get to $B_K^\msbar$ we need to renormalize $B_K^{\rm lat}$ and
control mixing with other dimension six, chirally disallowed
operators. Using NPR \cite{RBC-NPR} in the quenched theory we have 
seen that  mixing coefficients with chirally disallowed operators are small.
Preliminary NPR calculations on our $\sim 40$ dynamical DWF lattices
also show that the mixing coefficients are small (less than a few percent).  

These NPR results are preliminary, since we have not fully investigated all
the mass dependence of the procedure and our statistics are somewhat
limited.  However, we can also use the $\Omega$ parameter introduced in
\cite{RBC-WME} to quantify the 
chiral symmetry breaking of DWF and
conclude that the mixing with chirally disallowed operators is of
${\cal O}(\mres^2)$ and hence negligible.

We calculate $\hat{B}_K$, the renormalization group invariant (RGI)
parameter, from $B_K^{\rm lat}$ via the multiplicative renormalization
\begin{eqnarray}
  \hat{B}_K & = & Z^{\rm RGI,RI-MOM}(\mu) \nonumber \\
    & \times & ( Z_{Q}^{\Delta S=2}(\mu a)/Z_A^2(\mu a ) )
    B_K^{\rm lat}(a)
\end{eqnarray}
A similar equation is used for $B_K^\msbar$ except that $ Z^{\rm RGI,
\rm RI-MOM} \to Z^{\msbar, {\rm RI-MOM}}$.  The factor $ Z^{\rm
RGI,RI-MOM} \, Z_{Q}^{\Delta S=2}(\mu)/Z_A^2(\mu)$ is calculated
following the techniques in \cite{RBC-NPR} with all quark masses equal
to $\mdyn$.  
This quantity is extrapolated to $\mdyn = -\mres$ then
the extrapolation to $(pa)^2= 0$ is done to remove ${\cal O}(a^2)$ errors.
The two loop $\alpha_S(\mu)$ formula with $N_F=2,
\Lambda_{\rm QCD}=300$ MeV and $a^{-1}= 1.81 {\rm GeV}$ is used.  The
uncertainty due to the choice of $\Lambda_{\rm QCD}$ is estimated by
changing $\Lambda_{\rm QCD}$ by $\pm$50 MeV. 
We check the results
stay same within error by changing the order of
the two limits, $(pa)^2\to0$ and $\mdyn\to-\mres$.

Following the procedure above, we find $Z^{{\rm RGI}, {\rm RI-MOM}}
Z_{Q}^{\Delta S=2}/Z_A^2 = 1.29(4)$ and $Z^{\msbar, {\rm RI-MOM}}
Z_{Q}^{\Delta S=2}/Z_A^2  = 0.93(2)$ at $\mu=2$ GeV.
These numbers agree with the one loop perturbative calculation,
which gives the $\msbar$ factor to be 0.92\cite{DWF-pertZ}.
\section{CONCLUSIONS}

Our initial
calculation of $B_K$ with dynamical domain wall fermions has yielded
$B_K^{\msbar}(2 \; {\rm GeV}) = 0.503(20)$ and $ \hat{B}_K =
0.697(33)$ 
. 
The errors on these quantities include the statistical
error on $B_K^{\rm lat}$, the error on the $Z$ factors and a 2\% error
reflecting the uncertainty in the kaon mass arising from the
uncertainty in the lattice scale. 
These results correspond to an extrapolation of $\mdyn=\mval$ to 
$\mhalfstrangebar$.

We summarize results for each dynamical lattice in Table \ref{tab:B_K}.
These are obtained by interpolating $\mval$ to $\mhalfstrangebar$ at fixed
$\mdyn$ and are somewhat smaller than quenched DWF calculations at a similar
lattice spacing. 
We see a trend in Table \ref{tab:B_K} toward lower
values as the dynamical quark mass is reduced.  
We have likely underestimated our statistical error, 
due to the long autocorrelation
times in dynamical QCD simulations, and we have not yet studied 
the other methods of chiral extrapolations,
the continuum limit or finite volume effects.  However, it is clear that
these dynamical DWF calculations are possible with current computers
and hold promise for the next generation of computers.

\begin{table}
\caption{Preliminary results for renormalized $B_K$
from valence quark extrapolations.  For $\mdyn = 0.02,
0.03$ and 0.04 a 2\% scale error has been included in the error
on the renormalized quantities.  
}
\label{tab:B_K}
\begin{tabular}{llll}
\hline
$\mdyn$ & $B_K^{\rm lat}$ & $B_K^\msbar(2{\rm GeV})$ & $\hat{B}_K$ \\  
\hline
0.02 & 0.537(11) & 0.499(22) & 0.692(34) \\
0.03 & 0.557(9)  & 0.518(20) & 0.719(32) \\
0.04 & 0.568(10) & 0.529(21) & 0.733(34) \\
$\infty$ \cite{RBC-WME} &   & 0.536(6)  &\\
$\infty$ \cite{CPPACS-BK} & & 0.564(14) &\\
\hline 
\end{tabular}
\end{table}

\end{document}